\documentclass[a4paper, 11pt, notitlepage]{article}

\usepackage{amsmath}
\usepackage{amsthm}
\usepackage{amssymb}
\usepackage{delarray}
\usepackage[dvipdfmx]{graphicx}
\usepackage{color}
\usepackage{here}

\usepackage{subcaption}

\allowdisplaybreaks

\begin{document}

\begin{titlepage}

\vspace*{20mm}

\baselineskip 25pt   

\begin{center}
{\Large\bf
Expansion method for pricing foreign exchange options \\ under stochastic volatility and interest rates
}

\vspace{10mm}   

\renewcommand{\thefootnote}{\fnsymbol{footnote}}

Kenji~Nagami
\footnote[2]{Mitsubishi UFJ Trust Investment Technology Institute Co., Ltd. (MTEC), 4-2-6 Akasaka, \\    \hspace{6mm}Minato city, Tokyo 107-0052, Japan.}
\footnote[3]{The views expressed here are those of the author and do not represent the official views \\ \hspace{6mm}of the company.}

August 26, 2019

\renewcommand{\thefootnote}{\arabic{footnote}}

\vspace{20mm}   

\end{center}

\begin{abstract}
Some expansion methods have been proposed for approximately pricing options which has no exact closed formula. 
Benhamou et al. (2010) presents the smart expansion method that directly expands the expectation value of payoff function with respect to the volatility of volatility, then uses it to price options in the stochastic volatility model.
In this paper, we apply their method to the stochastic volatility model with stochastic interest rates, and present the expansion formula for pricing options up to the second order.
Then the numerical studies are performed to compare our approximation formula with the Monte-Carlo simulation.
It is found that our formula shows the numerically comparable results with the method proposed by Grzelak et al. (2012) which uses the approximation of characteristic function.
\end{abstract}

\end{titlepage}

\newpage
\baselineskip 16pt
%%%%%%%%%%%%%%%%%%%%%%%%%%%%%%%%%%%%%%%%%%%%%%%%%%%%%%%%%%%%%%%%%%%%%%%%%%%%%%%%%%%%%%%%%%%%%%%%%%%%%%%%%%%%%%%%%%%%%%%%%%
\section{Introduction}

The foreign exchange rate is expected to move depending on the interest rate of each currency.
The higher interest rate currency is expected to be exchanged to the lower interest rate currency at the lower exchange rate in a future than the spot exchange rate in the arbitrage free market.
It is plausible to consider the FX model with stochastic interest rates. In fact, it would be certainly needed to price a kind of derivatives including exotic ones with payoff depending on both FX rate and interest rates.
One of those models is the so-called Heston-Hull-White model that represents the FX rate by the Heston model  and incorporates the stochastic interest rates described by the Hull-White models.
This FX model is thought to have no closed formula for pricing plain vanilla options,
so it will be practically useful if a fast pricing formula is available.

In this paper, we consider the option pricing problem in the Heston-Hull-White model and present the approximation formula by using the smart expansion method.
This method is presented in \cite{BenhamouGobetMiri2010} to obtain the approximation formula in the time dependent Heston model with the deterministic interest rates.
The expected value of payoff function is directly expanded with respect to the volatility of volatility (vol-of-vol) and explicitly calculated by taking advantage of the Malliavin calculus.
The smart expansion method appears to follow the apparently different way from the asymptotic expansion method which expands the density function of underlying asset, 
but both methods are likely to share a substantial part of concept as expanding by vol-of-vol and using the Malliavin calculus.

The asymptotic expansion method is based on \cite{Watanabe1987}, \cite{Yoshida1992a}, \cite{KunitomoTakahashi2003} 
and has been developed by many authors including \cite{Yoshida1992b}, \cite{Takahashi1995}, \cite{Takahashi1999}.
It is naturally applied to the option pricing problem in the Heston model and one can find the derivation of the approximation formula up to the second order in \cite{TakahashiYamada2015}.
Other expansion methods applied for pricing options in the Heston model can be also found in some papers including \cite{Lewis2000}, \cite{Alos2012} and \cite{LorigPagliaraniPascucci2014}.

The Heston model has the closed form expression of characteristic function, by which one can exactly price plain vanilla options.
For the Heston-Hull-White model, the approximation formula of characteristic function is obtained in \cite{GrzelakOosterlee2012} and they apply it to pricing options.
Our proposed method is also compared with their method in the numerical studies.

This paper is organized as follows. 
Section 2 reviews the smart expansion method of \cite{BenhamouGobetMiri2010} that is used for pricing options in the Heston model with the deterministic interest rates.
In Section 3, we apply the expansion method to the Heston-Hull-White model and derive the approximation formula for pricing options up to the second order of vol-of-vol.
Section 4 shows the results of numerical studies where our proposed method is compared with the Mote-Carlo simulation as benchmark.
Section 5 gives the conclusion.
The appendices give the results of representative calculations needed to derive and evaluate our formula.

%%%%%%%%%%%%%%%%%%%%%%%%%%%%%%%%%%%%%%%%%%%%%%%%%%%%%%%%%%%%%%%%%%%%%%%%%%%%%%%%%%%%%%%%%%%%%%%%%%%%%%%%%%%%%%%%%%%%%%%%%%
\section{Case of deterministic interest rates}

In this section, we briefly review Benhamou et al. (2010)~\cite{BenhamouGobetMiri2010} that uses the expansion method to derive the approximation formula for pricing options in the Heston model under the deterministic interest rates.

The Heston model can be described with the FX spot rate observed at time $t$ denoted as $S_t^\epsilon$ and the stochastic variance $v_t^\epsilon$ that follows a CIR process.
\begin{align}
dS_t^\epsilon &= S_t^\epsilon ( r_{d} - r_{f} ) dt + S_t^\epsilon \sqrt{ v_t^\epsilon} dW_{st}^Q, \quad S_0^\epsilon = S_0,\\
dv_t^\epsilon &= k_v (\theta_v - v_t^\epsilon ) dt + \epsilon \gamma \sqrt{ v_t^\epsilon} dW_{vt}^Q, \quad v_0^\epsilon = v_0 . \label{HestonVolatilitySDE}
\end{align}

$W_{st}^Q, W_{vt}^Q$ are Brownian motions with correlation $\rho_{sv}$ on the domestic risk-neutral measure $Q$.
$r_d, r_f$ are the domestic and foreign interest rates respectively and assumed to be constant or deterministic.
The FX rate is measured as amount of domestic currency exchanged with a unit of foreign currency.

The stochastic differential equations are perturbed with a parameter $\epsilon \in [0, \,1]$ which accompanies with the volatility of volatility (vol-of-vol) $\gamma$ and is used to expand the option premium and specify the order of vol-of-vol.

The FX forward rate with maturity $T$ fixed is evaluated at $t$ as
\begin{equation}
F^\epsilon_t = S^\epsilon_t e^{ \int_t^T (r_d - r_f) ds },
\end{equation}
which converges with the spot rate at the maturity $T$.
The forward rate is martingale under $Q$.
\begin{align}
dF_t^\epsilon / F_t^\epsilon 
= \sqrt{ v^\epsilon_t } dW_{st}^Q .
\label{Eq_F_SDE_Heston}
\end{align}

The plain vanilla put option with the maturity $T$ and the strike $K$ and a unit notional in foreign currency is considered and 
the price in the domestic currency is evaluated  by using the expectation on $Q$.
\begin{equation}
PV(\epsilon) = D_d(T) E \left[  \left( K - F^\epsilon_T  \right)^+ \right], \hspace{5mm} D_d(T) = e^{ - \int_0^T r_d dt } .
\label{PutOptionPriceHeston}
\end{equation}

%%%%%%%%%%%%%%%%%%%%%%%%%%%%%%%%%%%%%%%%%%%%%%%%%%%%%%%%%%%%%%%%%%%%%%%%%%%%%%%%%%%%%%%%%%%%%%%%%%%%%%%%%%%%%

\vspace*{5mm}
If it is conditioned with the filtration ${\cal F}_v$ generated by the volatility $v^\epsilon_t$, 
the log forward rate at the maturity is normally distributed. 
The relevant parameters are explicitly written as

\begin{eqnarray}
x^\epsilon 
&\equiv & E \left[ \log F^\epsilon_T |  {\cal F}_v \right] + \frac{1}{2} Var \left[ \log F^\epsilon_T |  {\cal F}_v \right] \nonumber \\
&=& \log F_0 - \frac{1}{2} \int_0^T \rho_{sv}^2 v^\epsilon_t \, dt 
                                              + \int_0^T \rho_{sv} \sqrt{ v^\epsilon_t }    \, dW_{vt}, \\
y^\epsilon 
&\equiv & Var \left[ \log F^\epsilon_T |  {\cal F}_v \right] \nonumber \\
&=& \int_0^T (1-\rho_{sv}^2) v^\epsilon_t \, dt .
\end{eqnarray}

The equation~(\ref{PutOptionPriceHeston}) can be written by using the conditional expectation with ${\cal F}_v$, which is explicitly evaluated by the Black-Scholes formula.
\begin{equation}
PV(\epsilon) %= D_d(T) E^T \left[ \left( K - F^\epsilon_T  \right) \right] 
                 = D_d(T) E \left[ E \left[ \left( K - F^\epsilon_T  \right)^+ | {\cal F}_v \right] \right]
                 =          E \left[  BS( x^\epsilon, y^\epsilon ) \right] . \label{EqCondExpBS}
\end{equation}
\begin{equation}
BS(x, y) = D_d(T) ( K \Phi(-d_2) - e^x \Phi(-d_1) ), d_1 = \frac{x-\log K + y/2}{\sqrt{y}}, d_2 = d_1 - \sqrt{y},
\end{equation}
where $\Phi(x)=\int_{-\infty}^x \phi(t) dt, \phi(x)=\frac{e^{-\frac{x^2}{2}}}{\sqrt{2\pi}}$.
This satisfies the following formula, which is useful to arrange the expansion formula later on.
\begin{equation}
\frac{\partial BS(x, y)}{\partial y} = \frac{1}{2} \left( \frac{\partial^2 BS(x, y)}{\partial x^2} - \frac{\partial BS(x, y)}{\partial x} \right) .
\label{EqBSderiv}
\end{equation}

The option price can be expressed by expanding the BS term with respect to $\epsilon$ up to the second order as
\begin{align}
PV(\epsilon) 
&= E \left[ BS( x_{(0)}, y_{(0)} )  \right]  
+     E \left[ \frac{\partial}{\partial x} BS( x_{(0)}, y_{(0)} ) \left( \epsilon x_{(1)} + \frac{ \epsilon^2 }{2} x_{(2)} \right)  \right] \nonumber \\
&\quad + E \left[ \frac{\partial}{\partial y} BS( x_{(0)}, y_{(0)} ) \left( \epsilon y_{(1)} + \frac{ \epsilon^2 }{2} y_{(2)} \right)  \right]  \nonumber \\
&\quad + \frac{1}{2} E \left[ \frac{\partial^2}{\partial x^2} BS( x_{(0)}, y_{(0)} ) \epsilon^2 x_{(1)}^2  \right]
          + \frac{1}{2} E \left[ \frac{\partial^2}{\partial y^2} BS( x_{(0)}, y_{(0)} ) \epsilon^2 y_{(1)}^2  \right] \nonumber \\
&\quad +  E \left[ \frac{\partial^2}{\partial x \partial y} BS( x_{(0)}, y_{(0)} ) \epsilon^2 x_{(1)} y_{(1)}  \right] + o(\epsilon^3) , \label{Eq_PriceExpansion}
\end{align}
where the arguments of BS formula $x^\epsilon, \, y^\epsilon$ are expanded as
\begin{align}
x^\epsilon &= x_{(0)} + \epsilon x_{(1)} + \frac{ \epsilon^2 }{2} x_{(2)} + o(\epsilon^3), \label{expansionx} \\
y^\epsilon &= y_{(0)} + \epsilon y_{(1)} + \frac{ \epsilon^2 }{2} y_{(2)} + o(\epsilon^3). \label{expansiony}
\end{align}
The approximation formula of option price is obtained from (\ref{Eq_PriceExpansion}) by collecting all of the terms up to the second order and replacing $\epsilon$ with $1$.

%%%%%%%%%%%%%%%%%%%%%%%%%%%%%%%%%%%%%%%%%%%%%%%%%%%%%%%%%%%%%%%%%%%%%%%%%

\vspace*{5mm}
To specify the expansion coefficients, it is needed to expand the volatility $v^\epsilon_t$ with respect to $\epsilon$.
Assuming that the volatility is expanded as
\begin{equation}
v^\epsilon_t = v_{0,t} + \epsilon v_{1,t} + \frac{\epsilon^2}{2} v_{2,t} + \cdots,
\end{equation}
then equating each side of the equation~(\ref{HestonVolatilitySDE}) with the same order of $\epsilon$ produces SDEs for the expansion coefficients.
\begin{align}
dv_{0,t} &= k_v ( \theta_v - v_{0,t} ) dt, \hspace{5mm} v_{0,0} = v_0, \\
dv_{1,t} &= -k_v v_{1,t} dt + \gamma \sqrt{v_{0,t}} dW_{v,t}, \hspace{5mm} v_{1,0} = 0, \\
dv_{2,t} &= -k_v v_{2,t} dt + \gamma \frac{v_{1,t}}{ \sqrt{v_{0,t}} } dW_{v,t}, \hspace{5mm} v_{2,0} = 0.
\end{align}

These are solved to give the integral form representations.
\begin{align}
v_{0,t} &= \theta_v + (v_0 - \theta_v) e^{-k_v t}, \label{Eqv0t} \\
v_{1,t} &= \gamma e^{-k_v t} \int_0^t e^{k_v u} \sqrt{v_{0,u}}  dW_{v,u}, \label{Eqv1t} \\
v_{2,t} &= \gamma e^{-k_v t} \int_0^t e^{k_v u} \frac{v_{1,u}}{ \sqrt{v_{0,u}} } dW_{v,u}. \label{Eqv2t}
\end{align}

The expansion coefficients $x_{(i)}, \, y_{(i)}, \, i=0, 1,2$ are specified with these volatility coefficients.
Although \cite{BenhamouGobetMiri2010} can allow for time dependent parameters, we show the result with constant parameters and set $v_0=\theta_v$ hereafter.

The zeroth order term is evaluated as the limit of vol-of-vol $\gamma \to 0$ where the model is reduced to the BS model.
\begin{equation}
E \left[ BS( x_{(0)}, y_{(0)} ) \right] 
= D_d(T) E \left[ \left( K - F^0_T \right)^+   \right] = BS( x_0, y_0 ), \hspace{5mm} 
x_0 = \log F_0, \; y_0 = v_0 T .
\label{Eq_x0_y0}
\end{equation}
Note that the BS formula is expanded around the parameters $x_{(0)}, y_{(0)}$ inside the expectation, then the slightly different parameters $x_0, y_0$ are used after evaluating directly the unconditional expectation. 
This equation can be also extended for derivatives of BS formula.
\begin{align}
E \left[ \frac{\partial^{i+j}}{\partial x^i \partial y^j} BS( x_{(0)}, y_{(0)} ) \right]  = \frac{\partial^{i+j}}{\partial x^i \partial y^j} BS( x_0, y_0 ), \hspace{5mm} i,j = 0,1,\ldots .
\label{ExpBSderiv}
\end{align}

The higher order terms in the expansion formula (\ref{Eq_PriceExpansion}) can be explicitly calculated by taking advantage of the Malliavin calculus. Based on the lemma 1.2.1 in \cite{Nualart2006}, the following lemma for the Brownian motion $W_t$ is derived in \cite{BenhamouGobetMiri2010}.
\begin{align}
E \left[ G \left( \int_0^T g(t) dW_t \right) \int_0^T \mu_t dW_t \right] 
= E \left[ G^{(1)} \left( \int_0^T g(t) dW_t \right)  \int_0^T g(t) \mu_t dt  \right] , \label{LemmaMalliavin}
\end{align}
where $G$ is a smooth function and $g$ is a deterministic function and $\mu_t$ is a square integrable and predictable process.
In the current case, $G$ is identified with the BS formula or its derivatives and
the argument of $G$ corresponds to the stochastic integral appeared in $x_{(0)}$ with $g = \rho_{sv} \sqrt{v_0}$, 
so the derivative of $G$ is performed with respect to $x$.

Finally the approximation formula up to the second order of vol-of-vol is obtained.
\begin{align}
P_{approx}^H (x_0, y_0)
&= BS(x_0, y_0) + \rho_{sv} v_0 \gamma \partial_x \partial_y BS(x_0, y_0) \int_0^T dt e^{k_v t} \int_t^T du e^{-k_v u}  \notag\\
&\quad + \rho_{sv}^2 v_0 \gamma^2 \partial_x^2 \partial_y BS(x_0, y_0) \int_0^T dt e^{k_v t} \int_t^T du \int_u^T ds e^{-k_v s}  \notag\\
&\quad + v_0 \gamma^2 \partial_y^2 BS(x_0, y_0) \int_0^T dt e^{2k_v t} \int_t^T du e^{-k_v u}  \int_u^T ds e^{-k_v s}  \notag\\
&\quad + \frac{1}{2} \rho_{sv}^2 v_0^2 \gamma^2 \partial_x^2 \partial_y^2 BS(x_0, y_0)  \left( \int_0^T dt e^{k_v t} \int_t^T du e^{-k_v u}  \right)^2 .
\label{EqApproxHeston}
\end{align}

%%%%%%%%%%%%%%%%%%%%%%%%%%%%%%%%%%%%%%%%%%%%%%%%%%%%%%%%%%%%%%%%%%%%%%%%%%%%%%%%%%%%%%%%%%%%%%%%%%%%%%%%%%%%%%%%%%%%%%%%%%
\section{Heston-Hull-White model}

The previous section assumes the interest rates to be constant or deterministic.
In this section, we consider the FX model under the stochastic interest rates,
assuming that the domestic short rate $r_{dt}$ and the foreign short rate $r_{ft}$ follow the Hull-White models.
The FX spot rate $S_t^\epsilon$ follows the Heston model as before. So it is called the Heston-Hull-White model.

The stochastic differential equations are written under the domestic risk-neutral measure $Q$ as
\begin{eqnarray}
dS_t^\epsilon &=& S_t^\epsilon ( r_{dt} - r_{ft} ) dt + S_t^\epsilon \sqrt{ v_t^\epsilon} dW_{st}^Q, \quad S_0^\epsilon = S_0, \\
dv_t^\epsilon &=& k_v (\theta_v - v_t^\epsilon ) dt + \epsilon \gamma \sqrt{ v_t^\epsilon} dW_{vt}^Q, \quad v_0^\epsilon = v_0, \\
dr_{dt} &=& k_d (\theta_d - r_{dt} ) dt + \eta_{d} dW_{dt}^Q, \\
dr_{ft} &=& ( k_f (\theta_f - r_{ft} ) - \eta_f \rho_{Sf} \sqrt{ v_t^\epsilon } ) dt + \eta_{f} dW_{ft}^Q .
\end{eqnarray}

These equations are perturbed with a parameter $\epsilon \in [0, \,1]$ next to the volatility of volatility (vol-of-vol) $\gamma$, which is again used to expand the option premium.
The foreign interest rate has an additional drift term originating from the measure change from the foreign risk-neutral measure to the domestic one.
The parameters $\theta_d, \, \theta_f$ are deterministic functions of time and to be determined with the initial values $r_{d0}, \, r_{f0}$ by using the observed curves.

The FX forward rate with maturity $T$ fixed is evaluated at $t$ as
\begin{equation}
F^\epsilon_t = S^\epsilon_t \frac{P_f(t,T)}{P_d(t,T)},
\end{equation}
where the price of the domestic discount bond with maturity $T$ observed at $t$ is denoted by $P_d(t,T)$, and the foreign one is denoted as well.

As the previous section, the put option with the maturity $T$ and the strike $K$ and a unit notional in foreign currency is considered and priced in the domestic currency.
It is convenient to express the put option premium in terms with the forward rate and the domestic forward measure $Q^T$ 
which uses $P_d(t,T)$ as the num$\acute{\mbox{e}}$raire.
\begin{equation}
PV_{HHW}(\epsilon) = E \left[ e^{ - \int_0^T dt r_{dt} } \left( K - S^\epsilon_T  \right)^+ \right] = D_d(T) E^T \left[ \left( K - F^\epsilon_T  \right)^+ \right].
\label{PutOptionPrice}
\end{equation}
The expectation symbol with the superscript $T$ means to be evaluated under $Q^T$. \\
$D_d(T) = P_d(0,T)$ is the domestic discount bond price with maturity $T$ observed at $t=0$.

The forward rate is martingale under $Q^T$ as explicitly derived in \cite{GrzelakOosterlee2012}.
\begin{align}
dF_t^\epsilon / F_t^\epsilon 
= \sqrt{ v^\epsilon_t } dW_{st}^T - \eta_d B_d(t,T) dW_{dt}^T + \eta_f B_f(t,T) dW_{ft}^T 
\equiv \sigma_F (t, v^\epsilon_t) dW_{Ft}^T ,  \label{Eq_F_SDE}
\end{align}
where the deterministic functions specific to the Hull-White model are defined by
\begin{equation}
B_d(t,T) = \frac{1}{k_d} ( e^{-k_d(T-t)} - 1), \; B_f(t,T) = \frac{1}{k_f} ( e^{-k_f(T-t)} - 1).
\end{equation}

The variance of forward rate is explicitly written as
\begin{eqnarray}
\sigma_F (t, v^\epsilon_t )^2
&=& v^\epsilon_t + \eta_d^2 B_d(t,T)^2 + \eta_f^2 B_f(t,T)^2 -2\rho_{Sd} \eta_d B_d(t,T) \sqrt{v^\epsilon_t} \nonumber \\
&& +2\rho_{Sf} \eta_f B_f(t,T) \sqrt{v^\epsilon_t} -2\rho_{df} \eta_d \eta_f B_d(t,T) B_f(t,T).
\end{eqnarray}

If the correlation between forward rate and volatility is expressed as $dW_{Ft}^T dW_{vt}^T = \rho_{Fv} (t, v^\epsilon_t) dt$,  
it satisfies
\begin{equation}
\sigma_F (t, v^\epsilon_t) \rho_{Fv} (t, v^\epsilon_t) = \rho_{S v} \sqrt{v^\epsilon_t} - \eta_d \rho_{vd} B_d(t, T) + \eta_f \rho_{vf} B_f(t, T) .
\end{equation}

The equation (\ref{Eq_F_SDE}) includes no term that is explicitly dependent on the short rates.
If the interest rates are described by other models, the short rates may appear in the equation.

In the current case, the equation shows that the log forward rate at the maturity is normally distributed with the filtration ${\cal F}_v$ conditioned.
Assuming that the expectations are replaced with those under $Q^T$,
the option price is evaluated as in the equation (\ref{EqCondExpBS}) and expanded as in the formula (\ref{Eq_PriceExpansion}). The arguments of BS formula are expressed as
\begin{eqnarray}
x^\epsilon 
&\equiv & E^T \left[ \log F^\epsilon_T |  {\cal F}_v \right] + \frac{1}{2} Var^T \left[ \log F^\epsilon_T |  {\cal F}_v \right] \nonumber \\
&=& \log F_0 - \frac{1}{2} \int_0^T \sigma_F^2 (t, v^\epsilon_t) \rho_{Fv}^2(t, v^\epsilon_t) \, dt 
                                              + \int_0^T \sigma_F    (t, v^\epsilon_t) \rho_{Fv}(t, v^\epsilon_t)    \, dW_{vt}^T, \\
y^\epsilon 
&\equiv & Var^T \left[ \log F^\epsilon_T |  {\cal F}_v \right] \nonumber \\
&=& \int_0^T \sigma_F^2 (t, v^\epsilon_t) ( 1- \rho_{Fv}^2(t, v^\epsilon_t) ) \, dt .
\end{eqnarray}

The SDE of volatility is expressed under $Q^T$ as
\begin{equation}
dv^\epsilon_t = (k_v ( \theta_v - v^\epsilon_t ) + \epsilon \gamma \rho_{vd} \eta_d B_d(t,T) \sqrt{v^\epsilon_t} ) dt + \epsilon \gamma \sqrt{v^\epsilon_t} dW_{vt}^T, \hspace{5mm} v^\epsilon_0 = v_0,
\label{VolatilitySDE}
\end{equation}
where the additional drift term is appeared through the measure change from $Q$ to $Q^T$.
We assume no correlation between the FX volatility and the interest rates $\rho_{vd}=0, \; \rho_{vf}=0$ hereafter, since they appear to be redundant in the usual practical case.
The issue about their roles would remain as a future subject.
Then the volatility has the same expansion coefficients as in $Q$.
It is also assumed that $v_0 = \theta_v$ as before.

%%%%%%%%%%%%%%%%%%%%%%%%%%%%%%%%%%%%%%%%%%%%%%%%%%%%%%%%%%%%%%%%%%%%%%%%%%%%%%%%%%%%%%%%%%%%%%%%%%
The expansion coefficients as defined in the equations (\ref{expansionx}) (\ref{expansiony}) are explicitly written here.
\begin{eqnarray}
x_{(0)} &=& \log F_0 - \frac{\rho_{sv}^2}{2} v_0 T  
                                     +  \rho_{sv} \sqrt{v_0} \int_0^T \, dW_{v,t}^T,  \\
x_{(1)} &=& - \frac{\rho_{sv}^2}{2} \int_0^T v_{1,t} dt  + \frac{ \rho_{sv} }{ 2\sqrt{v_0} } \int_0^T v_{1,t} dW_{v,t}^T , \\
x_{(2)} &=& - \frac{\rho_{sv}^2}{2} \int_0^T v_{2,t} dt  + \frac{ \rho_{sv} }{ 2\sqrt{v_0} } \int_0^T v_{2,t} dW_{v,t}^T 
               - \frac{\rho_{sv}}{4 v_0^{3/2}} \int_0^T v_{1,t}^2 dW_{v,t}^T , \\ 
y_{(0)} &=& \int_0^T \sigma_F^2 (t,v_0) (1- \rho_{Fv}^2(t,v_0) ) dt, \\
y_{(1)} &=& \int_0^T \left( 1- \rho_{sv}^2 + \alpha(t)  \right) v_{1,t} dt \\
y_{(2)} &=& \int_0^T \left( 1- \rho_{sv}^2 + \alpha(t)  \right) v_{2,t} dt - \frac{1}{2 v_0} \int_0^T \alpha(t) v_{1,t}^2 dt, 
\end{eqnarray}
where we note that the additional terms specific to the stochastic interest rates are appeared,
\begin{equation}
\alpha(t) \equiv \frac{\rho_{sd} \eta_d}{\sqrt{v_0}} \frac{1-e^{-k_d(T-t)}}{k_d} 
                   - \frac{\rho_{sf} \eta_f}{\sqrt{v_0}} \frac{1-e^{-k_f(T-t)}}{k_f}.
\label{alpha}
\end{equation}

The zeroth order term is evaluated as the limit of vol-of-vol $\gamma \to 0$ where the model is reduced to the BS-Hull-White model.
\begin{equation}
E^T \left[ BS( x_{(0)}, y_{(0)} ) \right] 
= D_d(T) E^T \left[ \left( K - F^0_T \right)^+   \right] = BS( x_0, y_0 ), \; x_0 = \log F_0, \; y_0 = \int_0^T \sigma_F (t, v_0)^2 dt.  
\end{equation}
This equation corresponds to the extended formula of (\ref{Eq_x0_y0}) to allow for the stochastic interest rates and is evaluated by the expectation under $Q^T$.
It can be also extended to the cases for derivatives of BS formula as in the equation (\ref{ExpBSderiv}) with $y_0$ defined above.
The arguments of the BS formula may be abbreviated.

The variance integral of the forward rate is explicitly written as
\begin{align} 
\int_0^T \sigma_F (t, v_0)^2 dt
&= v_0 T 
+ \eta_d^2 \left( \frac{T}{k_d^2} - \frac{3}{2k_d^3} + \frac{2 e^{-k_d T}}{k_d^3} - \frac{ e^{-2k_dT} }{2k_d^3}  \right) 
+ \eta_f^2 \left( \frac{T}{k_f^2}  - \frac{3}{2k_f^3} + \frac{2 e^{-k_f T}}{k_f^3}  - \frac{ e^{-2k_fT} }{2k_f^3}  \right)   \notag \\
& - 2 \sqrt{v_0} \rho_{sd} \eta_d \left( - \frac{T}{k_d} + \frac{1- e^{-k_d T}}{k_d^2}  \right) 
+ 2 \sqrt{v_0} \rho_{sf} \eta_f \left( - \frac{T}{k_f} + \frac{1- e^{-k_f T}}{k_f^2}  \right) \notag \\
& \quad - 2 \rho_{df} \frac{\eta_d}{k_d} \frac{\eta_f}{k_f} \left( T + \frac{1- e^{-(k_d + k_f) T }}{k_d + k_f} - \frac{1- e^{-k_d T} }{k_d} - \frac{1- e^{-k_f T} }{k_f}  \right). 
\label{Eq_y_0}
\end{align}

In what follows, we show the expressions of higher order terms which are evaluated by invoking the brute force calculation and arranged by using the equation (\ref{EqBSderiv}).
The results for the representative terms are summarized in the appendix A.
The partial derivatives may be abbreviated as $\partial_x^n = \frac{\partial^n}{\partial x^n}, \partial_y^n = \frac{\partial^n}{\partial y^n}, \; n=1,2,\ldots$.
The integrals may be represented in the manner that the derivative symbol for variable of integration  is set next to the integral as $\int dx f(x) = \int f(x) dx$
, so that it would be easy to read the integration interval of each variable especially for multiple integrals.
The abbreviation $\beta(t) \equiv 1- \rho_{sv}^2 + \alpha(t) $ is also used.

% x_1
\begin{align}
E^T \left[ \partial_x BS x_{(1)} \right]  
=  \rho_{sv}^3 v_0 \gamma \partial_x \partial_y BS(x_0, y_0) \int_0^T dt e^{k_v t} \int_t^T du e^{-k_v u} .
\end{align}

% y_1
\begin{align}
E^T \left[ \partial_y BS y_{(1)} \right]  
= \rho_{sv} v_0 \gamma \partial_x \partial_y BS(x_0, y_0) \int_0^T dt e^{k_v t} \int_t^T du e^{-k_v u} (1-\rho_{sv}^2 + \alpha(u) ) .
\end{align}

% x_2
\begin{align}
E^T \left[ \partial_x BS x_{(2)} \right]  
&= \rho_{sv}^4 v_0 \gamma^2 \partial_x^2 \partial_y BS(x_0, y_0) \int_0^T dt e^{k_v t} \int_t^T du \int_u^T ds e^{-k_v s}  \notag \\ %\label{Eq_x2_1}
&\quad - \frac{\rho_{sv}^2}{4} \gamma^2 \partial_x^2 BS(x_0, y_0) \int_0^T dt e^{2k_v t} \int_t^T du e^{-2k_v u}  \notag \\ %\label{Eq_x2_2}
&\quad - \frac{\rho_{sv}^4}{2} v_0 \gamma^2 \partial_x^4 BS(x_0, y_0)  \int_0^T dt e^{k_v t} \int_t^T du e^{k_v u} \int_u^T ds e^{-2 k_v s} . \label{Eq_x2}
\end{align}

% y_2
\begin{align}
E^T \left[ \partial_y BS y_{(2)} \right] 
&= \rho_{sv}^2 v_0 \gamma^2 \partial_x^2 \partial_y BS(x_0, y_0) \int_0^T dt e^{k_v t} \int_t^T du \int_u^T ds e^{-k_v s} (1-\rho_{sv}^2 + \alpha(s) ) \notag
\\
& \quad
- \frac{1}{2} \gamma^2 \partial_y BS(x_0, y_0) \int_0^T dt e^{2k_v t} \int_t^T du e^{-2k_v u} \alpha(u)  \notag
\\
& \quad
- \rho_{sv}^2 v_0 \gamma^2 \partial_x^2 \partial_y BS(x_0, y_0)  \int_0^T dt e^{k_v t} \int_t^T du e^{k_v u} \int_u^T ds e^{-2 k_v s} \alpha(s) . 
\label{Eq_y2}
\end{align}

% x_1^2
\begin{align}
& E^T \left[ \partial_x^2 BS x_{(1)}^2 \right] \notag \\
&=  \rho_{sv}^6 v_0^2 \gamma^2 \partial_x^2 \partial_y^2 BS(x_0, y_0) 
      \left( \int_0^T dt e^{k_v t} \int_t^T du e^{-k_v u} \right)^2 \notag \\
&\quad + 2             \rho_{sv}^4 v_0    \gamma^2                \partial_y^2 BS(x_0, y_0) \int_0^T dt e^{2k_v t} \int_t^T du e^{-k_v u} \int_u^T ds e^{-k_v s} \notag \\
&\quad +                \rho_{sv}^4 v_0    \gamma^2 \partial_x^2 \partial_y   BS(x_0, y_0) \int_0^T dt e^{k_v t} \int_t^T du                \int_u^T ds e^{-k_v s} \notag \\
&\quad + \frac{1}{2}  \rho_{sv}^4 v_0    \gamma^2 \partial_x^4                BS(x_0, y_0)  \int_0^T dt e^{k_v t} \int_t^T du e^{k_v u} \int_u^T ds e^{-2 k_v s} \notag \\
&\quad + \frac{1}{4}  \rho_{sv}^2         \gamma^2 \partial_x^2                BS(x_0, y_0) \int_0^T dt e^{2k_v t} \int_t^T du e^{-2k_v u} . 
\label{Eq_x12}
\end{align}

% y_1^2
\begin{align}
& E^T \left[ \partial_y^2 BS y_{(1)}^2 \right] \notag\\
&=  \rho_{sv}^2 v_0^2 \gamma^2 \partial_x^2 \partial_y^2 BS(x_0, y_0) 
     \left( \int_0^T dt e^{k_v t} \int_t^T du e^{-k_v u} \beta(u) \right)^2 \notag \\
&\quad + 2 v_0 \gamma^2                                  \partial_y^2 BS(x_0, y_0) \int_0^T dt e^{2k_v t} \int_t^T du e^{-k_v u} \beta(u) \int_u^T ds e^{-k_v s} \beta(s) .  
\label{Eq_y12}
\end{align}

% x_1 y_1
\begin{align}
& E^T \left[ \partial_x \partial_y BS x_{(1)} y_{(1)} \right]  \nonumber\\
&=                         \rho_{sv}^4 v_0^2 \gamma^2 \partial_x^2 \partial_y^2 BS(x_0, y_0) 
              \left( \int_0^T dt e^{k_v t} \int_t^T du e^{-k_v u} \right) 
              \left( \int_0^T ds e^{k_v s} \int_s^T dr e^{-k_v r} \beta(r) \right)  \notag \\
&\quad +                 \rho_{sv}^2 v_0    \gamma^2                 \partial_y^2 BS(x_0, y_0) \int_0^T dt e^{2k_v t} \int_t^T du e^{-k_v u} \int_u^T ds e^{-k_v s} \beta(s)   \notag \\
&\quad +                 \rho_{sv}^2 v_0    \gamma^2                 \partial_y^2 BS(x_0, y_0) \int_0^T dt e^{2k_v t} \int_t^T du e^{-k_v u} \beta(u) \int_u^T ds e^{-k_v s}   \notag \\
&\quad +  \frac{1}{2}  \rho_{sv}^2 v_0    \gamma^2  \partial_x^2 \partial_y    BS(x_0, y_0) \int_0^T dt e^{k_v t} \int_t^T du  \int_u^T ds e^{-k_v s} \beta(s)  .  
\label{Eq_x1_y1}
\end{align}

These expressions are to be substituted into the right hand side of equation (\ref{Eq_PriceExpansion}) 
with expectations under $Q^T$ used and $\epsilon=1$. 
Then we can finally obtain the approximation formula to evaluate the price up to the second order of vol-of-vol.
This can be calculated by using the explicit expressions summarized in the appendix B.
\begin{align}
P_{approx}^{HHW} (x_0, y_0; \alpha)
&= BS(x_0, y_0) + \rho_{sv} v_0 \gamma \partial_x \partial_y BS(x_0, y_0) \int_0^T dt e^{k_v t} \int_t^T du e^{-k_v u} (1+\alpha(u)) \notag\\
&\quad + \rho_{sv}^2 v_0 \gamma^2 \partial_x^2 \partial_y BS(x_0, y_0) \int_0^T dt e^{k_v t} \int_t^T du \int_u^T ds e^{-k_v s} (1+\alpha(s)) \notag\\
&\quad + v_0 \gamma^2 \partial_y^2 BS(x_0, y_0) \int_0^T dt e^{2k_v t} \int_t^T du e^{-k_v u} (1+\alpha(u)) \int_u^T ds e^{-k_v s} (1+\alpha(s)) \notag\\
&\quad + \frac{1}{2} \rho_{sv}^2 v_0^2 \gamma^2 \partial_x^2 \partial_y^2 BS(x_0, y_0)  \left( \int_0^T dt e^{k_v t} \int_t^T du e^{-k_v u} (1+\alpha(u)) \right)^2 \notag\\
&\quad -  \frac{1}{2} \rho_{sv}^2 v_0 \gamma^2 \partial_x^2 \partial_y BS(x_0, y_0) \int_0^T dt e^{k_v t} \int_t^T du e^{k_v u}                \int_u^T ds e^{-2k_v s} \alpha(s) \notag\\
&\quad - \frac{1}{4}                       \gamma^2                \partial_y   BS(x_0, y_0)   \int_0^T dt e^{2k_v t} \int_t^T du e^{-2k_v u} \alpha(u) .
\label{EqApprox}
\end{align}

The effect of stochastic interest rates is incorporated in the $\alpha$ dependent terms and $y_0$ specified in the equation (\ref{Eq_y_0}).
If the interest rates are deterministic, we obtain the formula with $\alpha$ terms disappeared and $y_0$ replaced with $v_0 T$, 
which corresponds with the previous section's result, $P_{approx}^{HHW} (x_0, v_0 T; 0) = P_{approx}^H (x_0, v_0 T) $.

The pure effect of stochastic interest rates is determined by
\begin{align}
\Delta P_{approx} = P_{approx}^{HHW} (x_0, y_0; \alpha) - P_{approx}^H (x_0, v_0 T) .
\end{align}

If the interest rates are deterministic, the option price is accurately calculated by using the characteristic function of the Heston model and it is denoted as $P_{ChF}$.
Therefore we can present the other approximation formula that is defined by the hybrid of the expansion method and the characteristic function method.
\begin{align}
\tilde{P}_{approx}^{HHW} (x_0, y_0; \alpha) = P_{ChF} + P_{approx}^{HHW} (x_0, y_0; \alpha) - P_{approx}^H (x_0, v_0 T)
\label{EqApproxExpChF}
\end{align}
It also appears that the price of pure Heston model takes a role as control variate.

%%%%%%%%%%%%%%%%%%%%%%%%%%%%%%%%%%%%%%%%%%%%%%%%%%%%%%%%%%%%%%%%%%%%%%%%%%%%%%%%%%%%%%%%%%%%%%%%%%%%%%%%%%%%
\section{Numerical Experiment}
In this section, we study the accuracy of the approximation formula (\ref{EqApprox}) (\ref{EqApproxExpChF}) against the Monte-Carlo simulation using the QE scheme in \cite{Andersen2008}, which we use as benchmark.
We also compare the accuracy with another method presented in \cite{GrzelakOosterlee2011}, \cite{GrzelakOosterlee2012}, which uses the approximation of characteristic function.

The model parameters are set to be the hypothetical values in \cite{BenhamouGobetMiri2010}, \cite{GrzelakOosterlee2012}.
The interest rates model parameters are
\begin{equation}
\eta_d = 0.7\%, \quad \eta_f= 1.2\% , \quad k_d = 1\%, \quad k_f=5\% ,
\end{equation}
and zero rates are $0\%$ for each currency. 
We could also use any zero rates, which simply change the forward rate and the deterministic part of FX drift term in the MC simulation.

The FX model parameters are
\begin{equation}
v_0 = 0.05, \quad \gamma = 0.3, \quad k_v = 3 .
\end{equation}

The correlation matrix is
\begin{equation}
\left(
\begin{matrix}
1 & \rho_{sv} & \rho_{sd} & \rho_{sf} \\
   &  1         & \rho_{vd} &  \rho_{vf} \\
   &             & 1          &  \rho_{df} \\
   &             &             & 1
\end{matrix}
\right)
=
\left(
\begin{matrix}
1 & -0.4 & -0.15 & -0.15 \\
   &  1   &   0     &  0     \\
   &       &   1     &  0.25 \\
   &       &         & 1
\end{matrix}
\right),
\end{equation}
where we assume no correlation between the FX volatility and the interest rates $\rho_{vd}=\rho_{vf}=0$ as explained before, 
though \cite{GrzelakOosterlee2012} can approximately allow for finite correlations.

The Monte-Carlo simulation is performed with the number of scenarios $10^6$ and the time interval $0.05$ year.
The initial FX forward rate is $F_0 = 100$. 
We evaluate the put options with the maturities $T=1,3,5,7,10$ years and the unit notional in the foreign currency.
The strikes are based on \cite{Piterbarg2006} and given by
\begin{equation}
K_i (T) = F_0 \exp( 0.1\delta_i \sqrt{T}), \hspace{5mm} \{\delta_i \}_{i=1,\ldots,7}= \{-1.5, -1, -0.5, 0, 0.5, 1, 1.5\}. \label{strikes}
\end{equation}

The implied volatilities and the prices of put options are shown in Table~\ref{Table:iv} and Table~\ref{Table:pv} respectively.
It is found that the differences of implied volatilities from the Monte-Carlo simulation are a few basis points for each method (1bp = 0.01\%).
The expansion based methods have a comparable accuracy with the approximate characteristic function method.
The difference of implied volatility is appeared to grow for long maturity, but it is also found that the difference of price remains to be comparable with the standard error of the Monte-Carlo simulation.
It is shown in \cite{BenhamouGobetMiri2010} that the price error is estimated as $o(\gamma^3 T^2)$ for the expansion method with the deterministic interest rates,
so we can expect that the expansion method with the stochastic interest rates may have the equal or larger price error.
The approximate characteristic function method is obtained by replacing the non-affine $\sqrt{v}$ terms with their expectation values in the Kolmogorov backward equation, so it is likely to have a better accuracy for shorter maturity.

%%%%%%%%%%%%%%%%%%%%%%%
\begin{table} [H]
\begin{center}
\includegraphics[width=12cm]{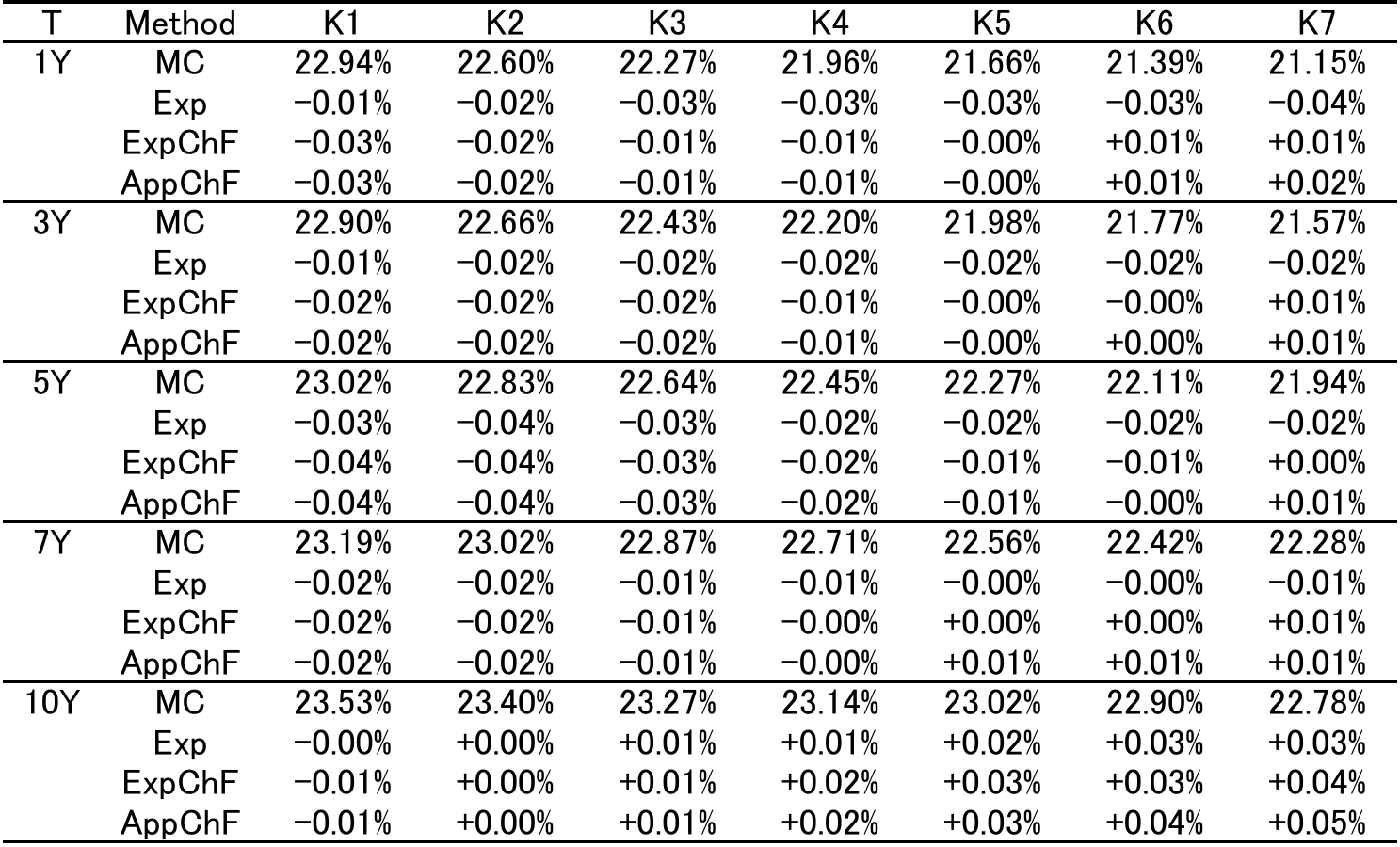}
\caption{ Implied volatilities of put options calculated by the Monte-Carlo simulation (MC) and the differences from MC for the expansion method (Exp),  the expansion \& characteristic function hybrid method (ExpChF) and the approximate characteristic function method (AppChF). Strikes $K_i, \, i=1,\ldots,7$ are given by (\ref{strikes}). }
\label{Table:iv}
\end{center}
\end{table}
%%%%%%%%%%%%%%%%%%%%%%%

%%%%%%%%%%%%%%%%%%%%%%%
\begin{table} [H]
\begin{center}
\includegraphics[width=12cm]{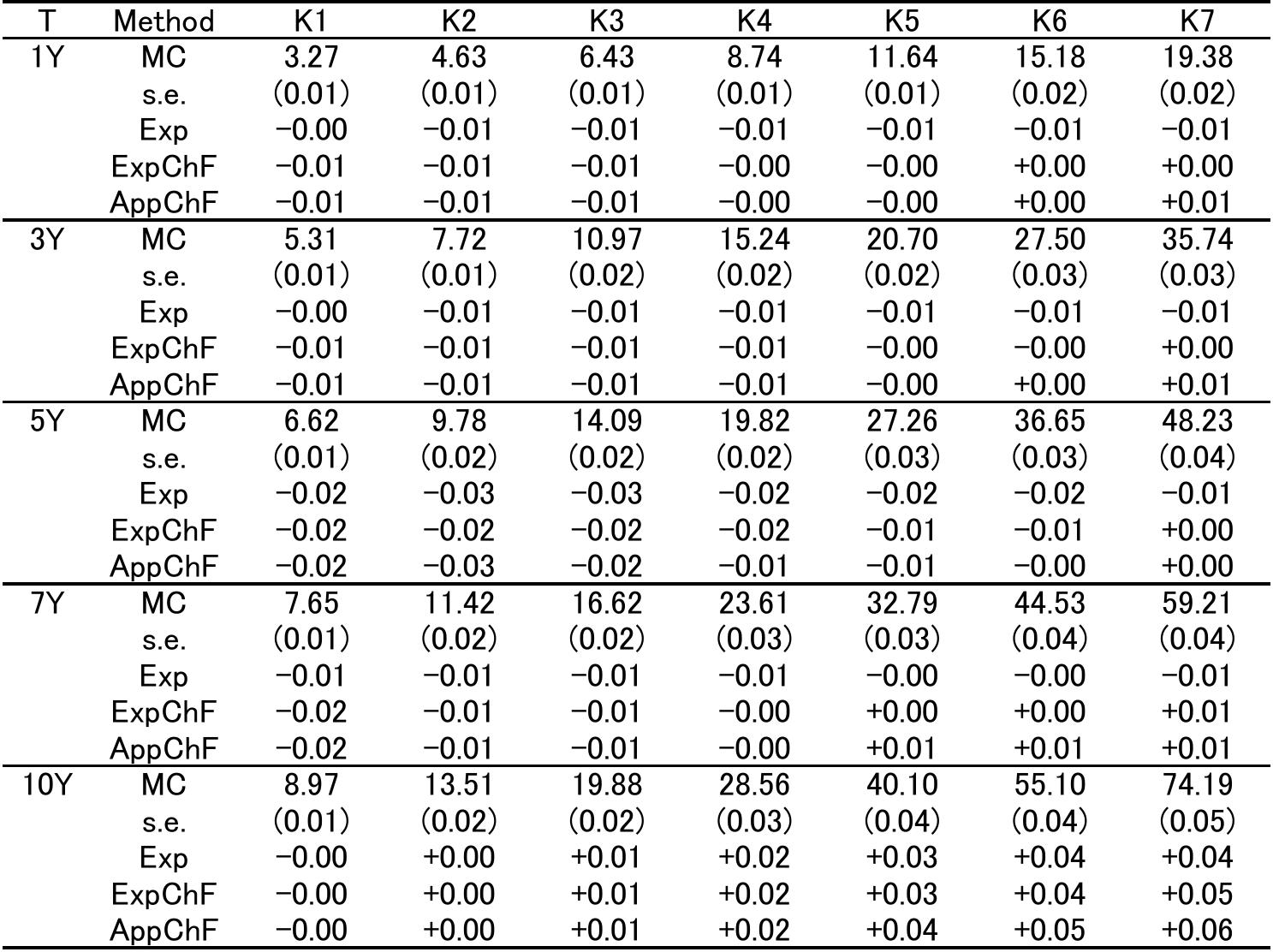}
\caption{ Prices of put options calculated by the Monte-Carlo simulation (MC) with the standard errors(s.e.),  the differences from MC for the expansion method (Exp),  the expansion \& characteristic function hybrid method (ExpChF) and the approximate characteristic function method (AppChF). }
\label{Table:pv}
\end{center}
\end{table}
%%%%%%%%%%%%%%%%%%%%%%%

%----------------------------
% Vol-of-Vol changed
%----------------------------
The expansion methods assume that the vol-of-vol is small enough to expand the option price with respect to it.
It is observed that the difference tends to grow with the vol-of-vol raised up in Table~\ref{Table:ivVolvol} where ATM options are used.
We note that the expansion \& characteristic function hybrid method has smaller difference than the pure expansion method.
The expansion \& ChF hybrid method partially takes advantage of the characteristic function method for the case of deterministic interest rates,
so it is plausible that the hybrid method has the higher accuracy than the pure expansion one for high vol-of-vol cases.

Note that the approximate characteristic function method is assumed to use for the low vol-of-vol cases.
The method replaces the non-affine $\sqrt{v}$ terms in PDE with the expectation value, which is evaluated by the approximation formula subject to $8 k_v \bar{v} \geq \gamma^2$ in \cite{GrzelakOosterlee2011}. Hence it is not simply applicable to the high vol-of-vol cases.
We would expect that the expansion \& ChF hybrid method can be practically used even for the somewhat high vol-of-vol cases.

%----------------------------
% ir volaility changed
%----------------------------
In Table~\ref{Table:ivSigma}, we report the cases with the interest rate volatility raised up.
The difference of implied volatility of options can be grown up as the interest rate volatility increases.
It is observed that the difference remains to be a few basis points if the interest rate volatility is a few percent.

As for the computational time, it takes 36 ms for the pure expansion method to evaluate 35 prices in Table~\ref{Table:pv}, 
while it takes 845 ms for the approximate characteristic function method. The Core i3 CPU 2.0GHz is used.
The pure expansion method is faster than the approximate characteristic function method by a factor 20.
The expansion \& ChF hybrid method takes 855 ms, which is comparable to the approximate characteristic function method.
The methods using the characteristic function need the numerical integration involving exponential functions and trigonometric functions, 
which dominates the computational cost.
The order of computational time for each method would be schematically written as 
\begin{align}
\mbox{(Heston/Exp)} < \mbox{(HHW/Exp)} \ll \mbox{(Heston/ChF)} < \mbox{(HHW/AppChF)} \sim \mbox{(HHW/ExpChF)}, \notag
\end{align}
where the left/right hand sides stand for the used model/method respectively.

%%%%%%%%%%%%%%%%%%%%%%%
\begin{table} [H]
\begin{center}
\includegraphics[width=15cm]{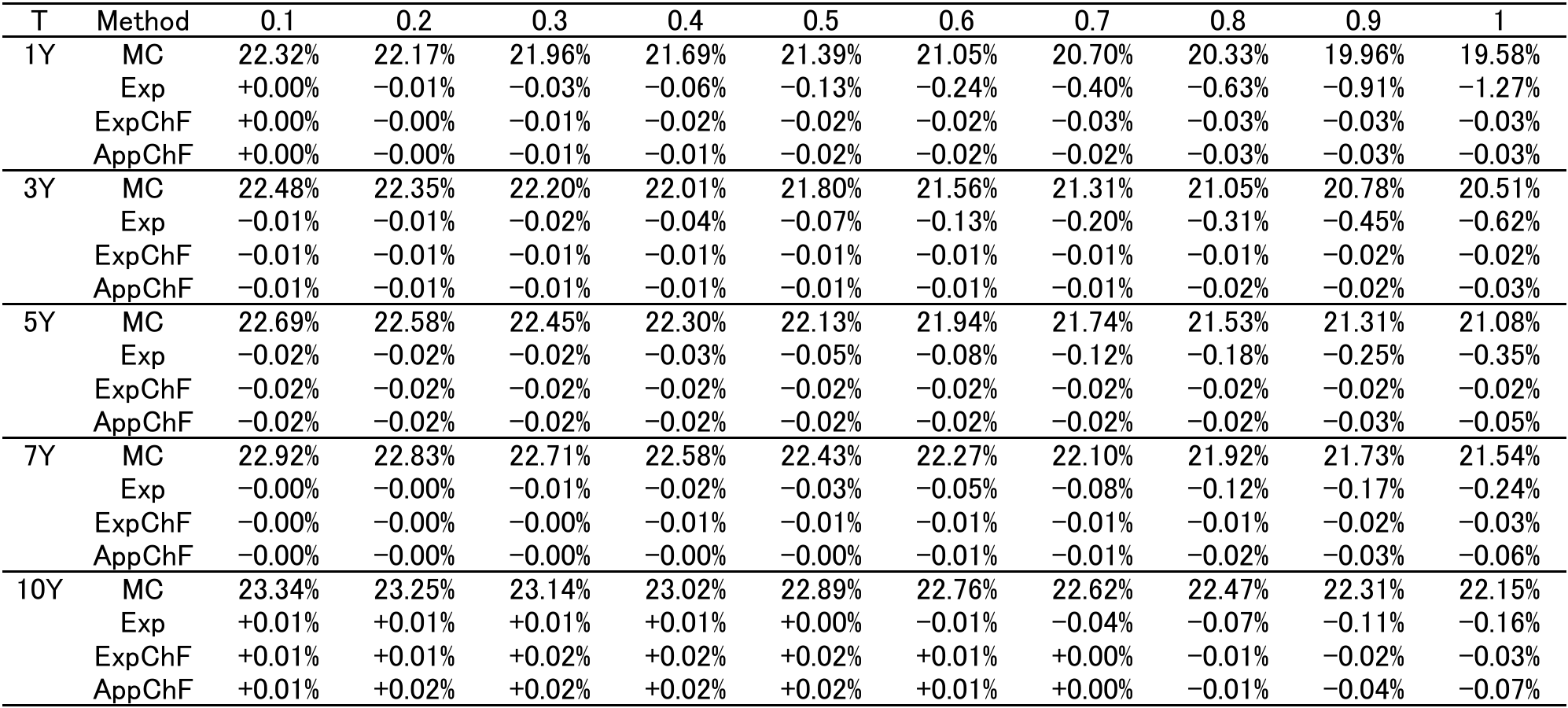}
\caption{ Implied volatilities of ATM put options with vol-of-vol specified on the top row and other parameters unchanged.  The Monte-Carlo simulation (MC) means the implied volatility of it and the differences from MC are displayed for the expansion method (Exp),  the expansion \& characteristic function hybrid method (ExpChF) and the approximate characteristic function method (AppChF).}
\label{Table:ivVolvol}
\end{center}
\end{table}
%%%%%%%%%%%%%%%%%%%%%%%

%%%%%%%%%%%%%%%%%%%%%%%
\begin{table}[H]
\begin{minipage}[t]{0.5\hsize}
\centering
\includegraphics[width=7.5cm]{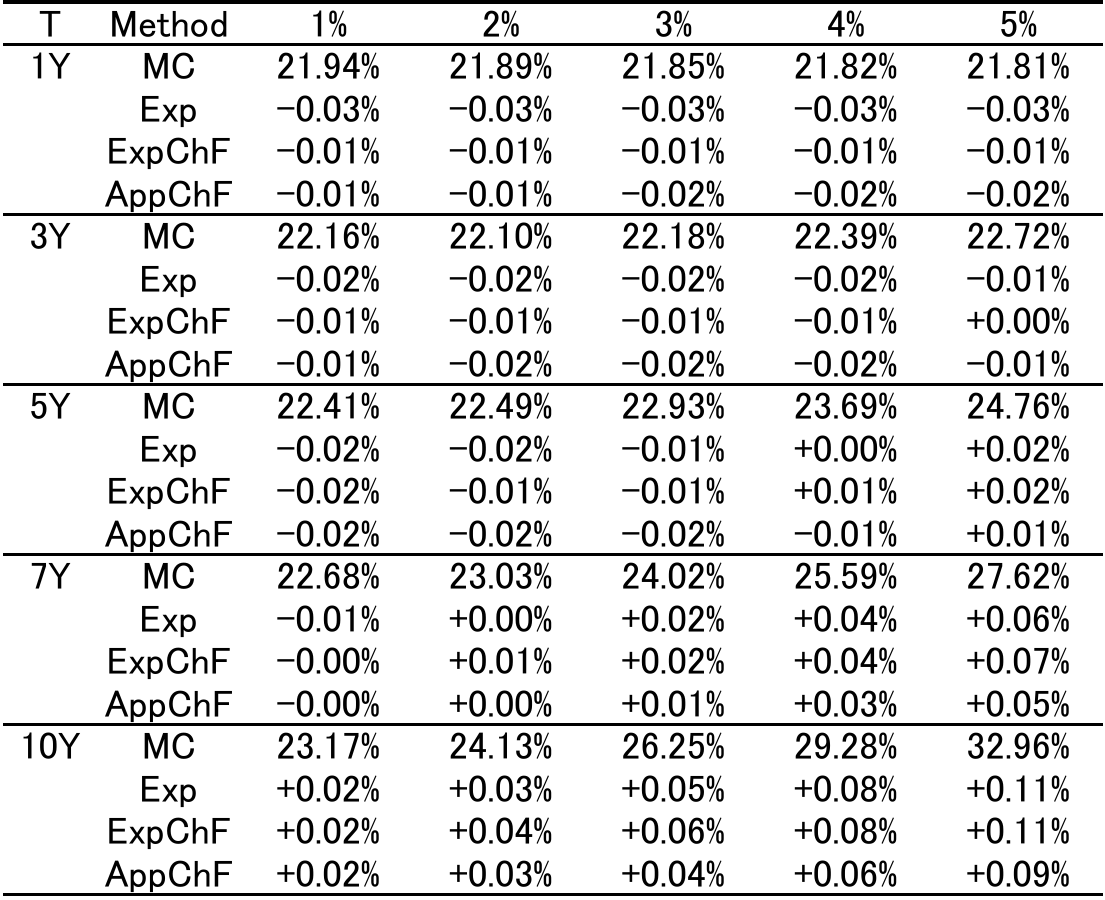}
\subcaption{domestic currency}
\end{minipage}
\begin{minipage}[t]{0.5\hsize}
\centering
\includegraphics[width=7.5cm]{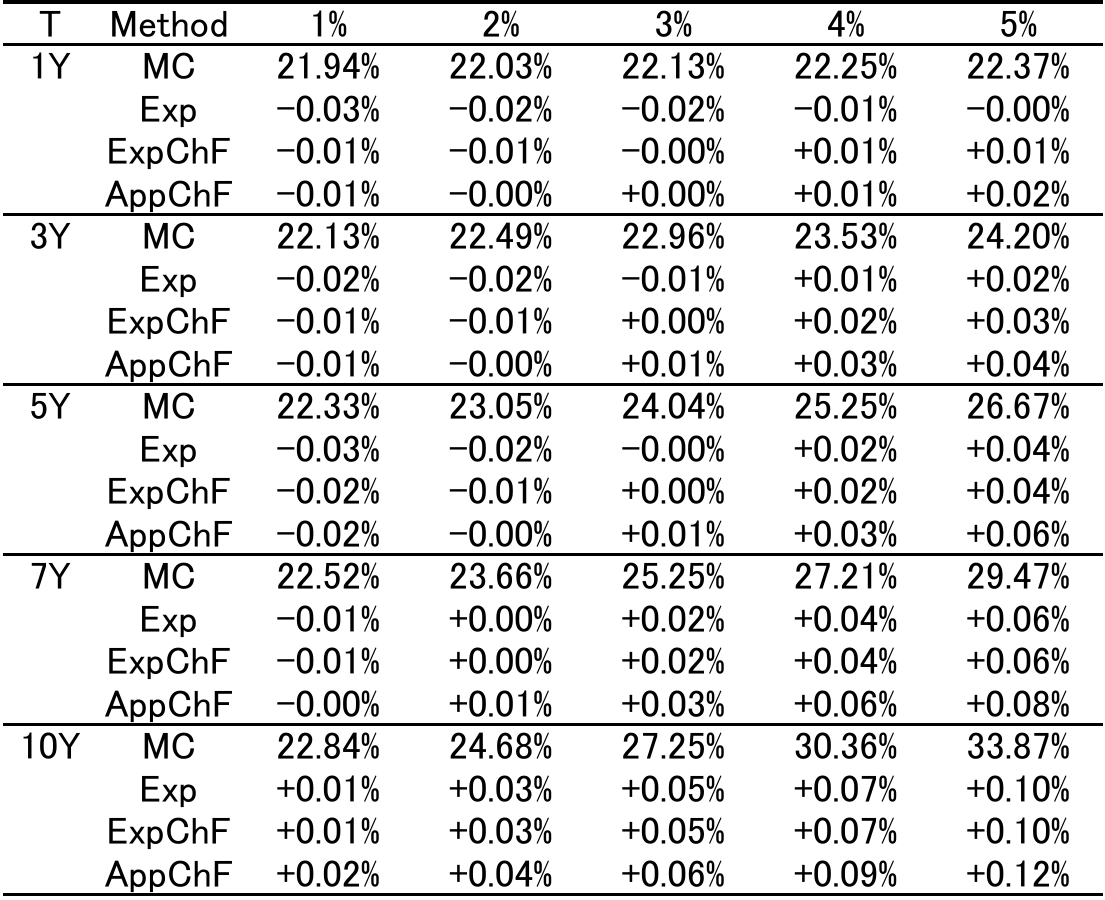}
\subcaption{foreign currency}
\end{minipage}
\caption{ Implied volatilities of ATM put options with interest rate volatility specified on the top row. The left table means varying domestic currency's interest rate volatility with foreign one and other parameters unchanged. The right table is the opposite case.}
\label{Table:ivSigma}
\end{table}
%%%%%%%%%%%%%%%%%%%%%%%

%%%%%%%%%%%%%%%%%%%%%%%%%%%%%%%%%%%%%%%%%%%%%%%%%%%%%%%%%%%%%%%%%%%%%%%%%%%%%%%%%%%%%%%%%%%%%%%%%%%%%%%%%%%%
\section{Conclusion}
As the foreign exchange rate model with the stochastic volatility and the stochastic interest rates, the Heston-Hull-White model is considered.
By using the expansion method of option premium with respect to the volatility of volatility (vol-of-vol),
we obtain the new approximation formula for pricing options up to the 2nd order in the Heston-Hull-White model.
It is inspired by \cite{BenhamouGobetMiri2010} in which the deterministic interest rates are used, and we partially extends it to the case with stochastic interest rates.
The approximation accuracy is numerically studied against the Monte-Carlo simulation.
If the volatility of volatility is not so high, the expansion based formula shows the comparable accuracy to the other method of ~\cite{GrzelakOosterlee2012},
which uses the approximation of characteristic function and is primarily applicable for low volatility of volatility.
In addition to the pure expansion method, we also present the hybrid method that incorporates the expansion method and the characteristic function method in \cite{Heston1993}.
In the numerical studies, the hybrid method shows the equivalent or higher accuracy than the pure expansion method and retains the good accuracy even for the high vol-of-vol cases.

\vskip 50pt
%%%%%%%%%%%%%%%%%%%%%%%%%%%%%%%%%%%%%%%%%%%%%%%%%%%%%%%%%%%%%%%%%%%%%%%%%%%%%%%%%%%%%%%%%%%%%%%%%%%%%%%%%%%%%%%%%%%%%%
\section*{Appendix A}

We summarize the results of representative calculations that is needed for evaluating the expansion terms.
They are obtained by using the It$\hat{\rm o}$ calculus and the lemma (\ref{LemmaMalliavin}) based on the Malliavin calculus.
As explained before, $G$ is a smooth function used to denote the BS formula or the derivatives of it,
and the derivatives of $G$ are equivalent to those with respect to $x$.
The expectation value of $G$ or its derivatives are explicitly evaluated by the equation (\ref{ExpBSderiv}).
$f$ is a deterministic function. Assuming $v_0 = \theta_v$ leads to $g = \rho_{sv} \sqrt{v_0}$.
$v_{i,t}, \, i=1,2$ are given by the equations (\ref{Eqv1t}), (\ref{Eqv2t}) respectively.
The right hand side is arranged by using the equality for some functions $f, g, h$,
\begin{align}
&\left( \int_0^T dt f(t) \int_t^T du g(u) \right) \left( \int_0^T ds f(s) \int_s^T dr h(r) \right) \notag\\
&=     \int_0^T dt f(t) \int_t^T du g(u) \int_u^T ds f(s) \int_s^T dr h(r) 
    +   \int_0^T dt f(t) \int_t^T du h(u) \int_u^T ds f(s) \int_s^T dr g(r) \notag\\
&\quad + 2 \left( \int_0^T dt f(t) \int_t^T du f(u) \int_u^T ds g(s) \int_s^T dr h(r)  
                     + \int_0^T dt f(t) \int_t^T du f(u) \int_u^T ds h(s) \int_s^T dr g(r) \right) . 
\label{Eq_MultipleIntegral}
\end{align}

In what follows, we express the equations under the risk neutral measure $Q$, 
but the same equations also hold under the $T$ forward measure $Q^T$ if the expectation and the Brownian motion are replaced with those defined under $Q^T$.

\begin{align}
&E \left[ G \left( \int_0^T g \, dW_{v,t} \right) \int_0^T f(t) v_{1,t} dt  \right] \nonumber \\
&= E \left[ G^{(1)} \left( \int_0^T g \, dW_v(t) \right) \right] \gamma \rho_{sv} v_0 \int_0^T dt e^{k_v t} \int_t^T du f(u) e^{-k_v u} .
\end{align}

\begin{align}
&E \left[ G \left( \int_0^T g \, dW_{v,t} \right) \int_0^T f(t) v_{1,t} dW_{v,t}  \right] \nonumber \\
&= E \left[ G^{(2)} \left( \int_0^T g \, dW_{v,t} \right) \right] \gamma \rho_{sv}^2 v_0^{3/2} \int_0^T dt e^{k_v t} \int_t^T du f(u) e^{-k_v u} .
\end{align}

\begin{align}
&E \left[ G \left( \int_0^T g \, dW_{v,t} \right) \int_0^T f(t) v_{2,t} dt  \right] \nonumber \\
& = E\left[ G^{(2)} \left( \int_0^T g \, dW_{v,t} \right) \right] \gamma^2 \rho_{sv}^2 v_0 \int_0^T ds e^{k_v s} \int_s^T du \int_u^T dt f(t) e^{-k_v t}.
\end{align}

\begin{align}
&E \left[ G \left( \int_0^T g \, dW_{v,t} \right) \int_0^T f(t) (v_{1,t})^2 dt  \right] \nonumber \\
& = E\left[ G \left( \int_0^T g \, dW_{v,t} \right) \right] \gamma^2 v_0 \int_0^T du \, e^{2k_v u} \int_u^T dt f(t) e^{-2k_v t} \nonumber \\
& \quad + 2 E\left[ G^{(2)} \left( \int_0^T g \, dW_{v,t} \right) \right]  \gamma^2 \rho_{sv}^2 v_0^2 \int_0^T ds e^{k_v s} \int_s^T du  e^{k_v u} \int_u^T dt f(t) e^{-2k_v t}. 
\end{align}

\begin{align}
& E \left[ G \left( \int_0^T g \, dW_{v,t} \right) \int_0^T (v_{1,t})^2 dW_{v,t}  \right] \nonumber \\
& = E\left[ G^{(1)} \left( \int_0^T g \, dW_{v,t} \right) \right]  \gamma^2 \rho_{sv} v_0^{3/2} \int_0^T dt  e^{2k_v t} \int_t^T du e^{-2k_v u}  \nonumber \\
& \quad + 2 E\left[ G^{(3)} \left( \int_0^T g \, dW_{v,t} \right) \right]  \gamma^2 \rho_{sv}^3 v_0^{5/2} \int_0^T ds  e^{k_v s} \int_s^T du  e^{k_v u} \int_u^T dt  e^{-2k_v t}.
\end{align}

\begin{align}
& E \left[ G \left( \int_0^T g \, dW_{v,t} \right) \int_0^T v_{2,t} dW_{v,t}  \right] \nonumber \\
& = E\left[ G^{(3)} \left( \int_0^T g \, dW_{v,t} \right) \right] \gamma^2 \rho_{sv}^3 v_0^{3/2} \int_0^T ds e^{k_v s} \int_s^T du  \int_u^T dt  e^{-k_v t}.
\end{align}

\begin{align}
&  E \left[ G\left( \int_0^T g \, dW_{v,t} \right) \int_0^T dt f_1(t) v_{1,t} \int_0^T du f_2(u) v_{1,u}  \right] \nonumber \\
& = E \left[ G^{(2)}\left( \int_0^T g \, dW_{v,t} \right) \right] \gamma^2 \rho_{sv}^2 v_0^2 
      \left( \int_0^T dt  e^{k_v t} \int_t^T dr f_1(r) e^{-k_v r} \right) 
      \left( \int_0^T ds  e^{k_v s} \int_s^T du f_2(u) e^{-k_v u} \right) \nonumber \\
& \quad +E \left[ G \left( \int_0^T g \, dW_{v,t} \right) \right] \gamma^2 v_0 \int_0^T ds e^{2k_v s} \int_s^T dt f_1(t) e^{-k_v t} \int_t^T du f_2(u) e^{-k_v u} \notag \\
& \quad +E \left[ G \left( \int_0^T g \, dW_{v,t} \right) \right] \gamma^2 v_0 \int_0^T ds e^{2k_v s} \int_s^T dt f_2(t) e^{-k_v t} \int_t^T du f_1(u) e^{-k_v u} .
\end{align}

\begin{align}
&  E \left[ G\left( \int_0^T g \, dW_{v,t} \right)  \int_0^T dW_{v,t} v_{1,t} \int_0^t dW_{v,s} v_{1,s} \right] \nonumber \\
& = \frac{1}{2} E \left[ G^{(4)}\left( \int_0^T g \, dW_{v,t} \right) \right] \gamma^2 \rho_{sv}^4 v_0^3 
     \left( \int_0^T dr  e^{k_v r} \int_r^T ds  e^{-k_v s} \right)^2 \nonumber \\
& \quad + E \left[ G^{(2)}\left( \int_0^T g \, dW_{v,t} \right) \right] \gamma^2 \rho_{sv}^2 v_0^2 \int_0^T ds e^{2k_v s} \int_s^T dt  e^{-k_v t} \int_t^T du  e^{-k_v u}  \nonumber \\
& \quad + E \left[ G^{(2)}\left( \int_0^T g \, dW_{v,t} \right) \right] \gamma^2 \rho_{sv}^2 v_0^2 \int_0^T dt  e^{k_v t} \int_t^T du \int_u^T ds  e^{-k_v s}.
\end{align}

\begin{align}
&  E \left[ G\left( \int_0^T g \, dW_{v,t} \right)  \left( \int_0^T f(t) v_{1,t}  dt \right) \left( \int_0^T v_{1,s} dW_{v,s} \right) \right] \nonumber \\
& =         E \left[ G^{(1)}\left( \int_0^T g \, dW_{v,t} \right) \right] \gamma^2 \rho_{sv} v_0^{3/2} \int_0^T ds e^{2k_v s} \int_s^T dt  e^{-k_v t} \int_t^T dr f(r) e^{-k_v r} \nonumber \\
& \quad + E \left[ G^{(1)}\left( \int_0^T g \, dW_{v,t} \right) \right] \gamma^2 \rho_{sv} v_0^{3/2} \int_0^T ds  e^{k_v s} \int_s^T dt \int_t^T dr f(r) e^{-k_v r} \nonumber \\
& \quad + E \left[ G^{(1)}\left( \int_0^T g \, dW_{v,t} \right) \right] \gamma^2 \rho_{sv} v_0^{3/2} \int_0^T dr e^{2k_v r} \int_r^T dt f(t) e^{-k_v t} \int_t^T ds  e^{-k_v s} \notag\\
& \quad + E \left[ G^{(3)}\left( \int_0^T g \, dW_{v,t} \right) \right] \gamma^2 \rho_{sv}^3 v_0^{5/2} \left( \int_0^T du  e^{k_v u} \int_u^T ds  e^{-k_v s} \right) \left( \int_0^T dt  e^{k_v t} \int_t^T dr f(r) e^{-k_v r} \right) .
\end{align}

\vskip 50pt
%%%%%%%%%%%%%%%%%%%%%%%%%%%%%%%%%%%%%%%%%%%%%%%%%%%%%%%%%%%%%%%%%%%%%%%%%%%%%%%%%%%%%%%%%%%%%%%%%%%%%%%%%%%%%%%%%%%%%%
\section*{Appendix B}
For completeness, we show the explicit formulas to evaluate the equation (\ref{EqApprox}).

Let the maturity $T$ fixed and write exponential terms as
\begin{align}
x_d = e^{k_d T}, 
\quad
x_f = e^{k_f T},
\quad
x_v = e^{k_v T},
\end{align}
and a function $c(t)$ as
\begin{align}
c(t) = c_0 + c_d e^{k_d t} + c_f e^{k_f t}.
\end{align}
If $c(t)$ is $\alpha(t)$ in (\ref{alpha}), the coefficients are given by
\begin{align}
c_0 = \frac{1}{\sqrt{v_0}} \left( \frac{\rho_{sd}\eta_d}{k_d} - \frac{\rho_{sf}\eta_f}{k_f} \right), 
\quad
c_d = - \frac{\rho_{sd}\eta_d}{k_d\sqrt{v_0} x_d}
\quad
c_f =   \frac{\rho_{sf}\eta_f}{k_f\sqrt{v_0} x_f}.
\end{align}
If $c(t) = 1+ \alpha(t)$, $c_0$ has an additional term $1$.

\begin{align}
\int_0^T dt e^{k_v t} \int_t^T du e^{-k_v u} c(u)
&= \frac{c_0 T}{k_v} + \frac{c_d (x_d -1)}{k_v k_d} + \frac{c_f (x_f -1)}{k_v k_f} + \frac{c_0 (1/x_v -1)}{k_v^2} \notag \\
&\quad - \frac{c_d (x_d/x_v -1)}{k_v(k_d - k_v)} - \frac{c_f (x_f/x_v -1)}{k_v(k_f - k_v)}
\end{align}
The similar integral $\int_0^T dt e^{2k_v t} \int_t^T du e^{-2k_v u} c(u)$ is also evaluated by replacing $k_v \to 2k_v$ and $x_v \to x_v^2$ in the above expression.

\begin{align}
&\quad \int_0^T dt e^{k_v t} \int_t^T du \int_u^T ds e^{-k_v s} c(s) \notag \\
&= c_0 \left( \frac{ T}{k_v^2} + \frac{1/x_v -1}{k_v^3} - \frac{ 1 - (1+k_v T)/x_v}{k_v^3} \right) \notag \\
&\quad 
+ c_d \left( \frac{x_d -1}{k_v^2 k_d} - \frac{x_d/x_v -1}{k_v^2 (k_d - k_v)} - \frac{1-(1-(k_d-k_v)T)x_d/x_v}{k_v (k_d-k_v)^2} \right) \notag \\
&\quad 
+ c_f \left( \frac{x_f -1}{k_v^2 k_f} - \frac{x_f/x_v -1}{k_v^2 (k_f - k_v)} - \frac{1-(1-(k_f-k_v)T)x_f/x_v}{k_v (k_f-k_v)^2} \right)
\end{align}

\begin{align}
&\quad \int_0^T dt e^{2k_v t} \int_t^T du e^{-k_v u} c(u) \int_u^T ds e^{-k_v s} c(s) \notag \\
&= c_0^2 \left( \frac{T}{2k_v^2} + \frac{1}{k_v^3 x_v} - \frac{1}{4k_v^3 x_v^2} - \frac{3}{4k_v^3} \right) \notag \\
&\quad 
+ c_d^2 \left( \frac{x_d/x_v -1}{(k_d+k_v)(k_d-k_v)^2} + \frac{x_d^2 -1}{4k_dk_v(k_d+k_v)} - \frac{x_d^2/x_v^2 -1}{4k_v(k_d-k_v)^2} \right) \notag \\
&\quad
+ c_f^2 \left( \frac{x_f/x_v -1}{(k_f+k_v)(k_f-k_v)^2} + \frac{x_f^2 -1}{4k_fk_v(k_f+k_v)} - \frac{x_f^2/x_v^2 -1}{4k_v(k_f-k_v)^2} \right) \notag \\
&\quad
+ c_0c_d \left( \frac{(k_d+2k_v)(x_d-1)}{2k_v^2(k_d+k_v)k_d} + \frac{x_d/x_v^2 -1}{2k_v^2(k_d-k_v)} + \frac{1-1/x_v}{k_v(k_d+k_v)(k_d-k_v)} 
- \frac{x_d/x_v -1}{k_v^2(k_d-k_v)} \right) \notag \\
&\quad
+ c_0c_f \left( \frac{(k_f+2k_v)(x_f-1)}{2k_v^2(k_f+k_v)k_f} + \frac{x_f/x_v^2 -1}{2k_v^2(k_f-k_v)} + \frac{1-1/x_v}{k_v(k_f+k_v)(k_f-k_v)} 
- \frac{x_f/x_v -1}{k_v^2(k_f-k_v)} \right) \notag \\
&\quad
+ c_dc_f \left( \frac{x_d/x_v-1}{(k_d-k_v)(k_f^2-k_v^2)} + \frac{x_f/x_v-1}{(k_f-k_v)(k_d^2-k_v^2)} - \frac{x_dx_f/x_v^2 -1}{2k_v(k_d-k_v)(k_f-k_v)} \right. \notag \\
&\hspace{25mm}
\left. + \left( \frac{1}{k_d+k_v} + \frac{1}{k_f+k_v} \right) \frac{x_dx_f -1}{2k_v(k_d+k_f)} \right)
\end{align}

\begin{align}
&\quad \int_0^T dt e^{k_v t} \int_t^T du e^{k_v u} \int_u^T ds e^{-2k_v s} c(s) \notag \\
&= c_0 \left( \frac{T}{2k_v^2} + \frac{1}{k_v^3 x_v} - \frac{1}{4k_v^3 x_v^2} - \frac{3}{4k_v^3} \right) \notag \\
&\quad
+ c_d \left( \frac{x_d -1}{2k_v^2 k_d} - \frac{x_d/x_v -1}{k_v^2 (k_d-k_v)} + \frac{x_d/x_v^2 -1}{2k_v^2(k_d-2k_v)} \right) \notag \\
&\quad
+ c_f \left( \frac{x_f -1}{2k_v^2 k_f} - \frac{x_f/x_v -1}{k_v^2 (k_f-k_v)} + \frac{x_f/x_v^2 -1}{2k_v^2(k_f-2k_v)} \right) 
\end{align}

Note that the above formulas should be modified with the limit value at the singular point.

The following expressions are to evaluate the derivatives of Black-Scholes formula. 
\begin{equation}
BS(x, y) = D_d(T) \left( K \Phi(-d_2) - e^x \Phi(-d_1) \right), d_1 = \frac{x-\log K + y/2}{\sqrt{y}}, d_2 = d_1 - \sqrt{y}
\end{equation}

\begin{align}
\frac{\partial BS(x, y)}{\partial y} = D_d(T) K \phi(d_2) \frac{1}{2\sqrt{y}}
\end{align}

\begin{align}
\frac{\partial^2 BS(x, y)}{\partial x \partial y} 
&= - D_d(T) K \phi(d_2) \frac{d_2}{2y}
\end{align}

\begin{align}
\frac{\partial^2 BS(x, y)}{\partial y^2} 
&= D_d(T) K \phi(d_2)  \frac{d_1d_2 - 1}{4y^{3/2}} 
\end{align}

\begin{align}
\frac{\partial^3 BS(x, y)}{\partial x^2 \partial y} 
&= D_d(T) K \phi(d_2) \frac{d_2^2 -1}{2y^{3/2}}
\end{align}

\begin{align}
\frac{\partial^4 BS(x, y)}{\partial x^2 \partial y^2} 
&= D_d(T) K \phi(d_2) \frac{d_1d_2^3 -3(d_1d_2+d_2^2 -1)}{4y^{5/2}}
\end{align}

\vskip 50pt
%%%%%%%%%%%%%%%%%%%%%%%%%%%%%%%%%%%%%%%%%%%%%%%%%%%%%%%%%%%%%%%%%%%%%%%%%%%%%%%%%%%%%%%%%%%%%%%%%%%%%%%%%%%%%%%%%%%%%%%%%%

\end{document}